\documentclass[a4paper,11pt]{article}
\usepackage[utf8]{inputenc}
\usepackage{pos}
\usepackage{support-caption} 
\usepackage{journals}

\usepackage{amsmath}

\RequirePackage[numbers,sort&compress]{natbib}

\newcommand{\GeV}{{\rm \,GeV}}

\newcommand{\MeV}{{\rm \,MeV}}

\newcommand{\cm}{{\rm \,cm}}
\newcommand{\km}{{\rm \,km}}
\newcommand{\s}{{\rm \,s}}

\newcommand{\erf}{{\rm \,Erf}}

 \def\be   {\begin{equation}}   \def\ee   {\end{equation}}
 \def\ba   {\begin{array}}      \def\ea   {\end{array}}
 \def\bea  {\begin{eqnarray}}   \def\eea  {\end{eqnarray}}
 \def\bean {\begin{eqnarray*}}  \def\eean {\end{eqnarray*}}

\newcommand{\Rstar}{R_\star}
\newcommand{\vstar}{v_\star}

\newcommand{\fFD}{f_{\rm FD}}

\newcommand{\m}{{\rm \,m}}

\newcommand{\sigmath}{\sigma_{th}}

\title{Capture of Dark Matter in Compact Stars}

\author*[a]{Giorgio Busoni}

\affiliation[a]{Max-Planck-Institut fur Kernphysik,\\ Saupfercheckweg 1, 69117 Heidelberg, Germany}

\emailAdd{giorgio.busoni@mpi-hd.mpg.de}

\abstract{
The extreme conditions in Neutron Stars make them ideal test facilities for fundamental interactions.
A Neutron Star can capture Dark Matter via scattering. As a result of the scattering, Dark Matter kinetic energy is transferred to the star.
An observational consequence of this can be the warming of old neutron stars to near-infrared temperatures.
Different approximations or simplifications have been applied to previous analyses of the capture process.
In this article, we present a significantly improved treatment of Dark Matter capture, which properly accounts for all relevant physical effects over a wide range of Dark Matter masses.
Among them are gravitational focusing, a fully relativistic scattering treatment, Pauli blocking, neutron star opacity and multiple scattering effects.
This paper provides general expressions that allow the capture rate to be computed numerically, and derives simplified expressions for particular types of interactions or mass regimes, which greatly increase the efficiency of computation.
As a result of our method, we are able to model the scattering of Dark Matter from any neutron star constituent as well as the capture of Dark Matter in other compact objects.
Our results are applied to scattering of Dark Matter from neutrons, protons, leptons and exotic baryons.
For leptonic targets, we find that a relativistic description is essential. In our analysis of the capture of Dark Matter in Neutron Stars, we include two important effects that are generally ignored by most studies.
Because the scattering of Dark Matter with nucleons in the star exhibits large momentum transfers, the nucleon structure must be considered via momentum-dependent hadronic form factors.
Moreover, because of the extreme densities of matter inside Neutron Stars, we should consider nucleon interactions instead of assuming all nucleons are a perfect Fermi gas.
Taking into account these effects results in a decrease of up to three orders of magnitude in the dark matter capture rate.

The potential Neutron Star sensitivity to DM-lepton scattering cross sections is much greater than electron-recoil experiments, particularly in the sub-GeV regime, with a sensitivity to sub-MeV DM well beyond the reach of future terrestrial experiments.
We also present results for DM-Baryon scatterings in Neutron Stars, where the sensitivity is expected to exceed that of current DD experiments for spin-dependent cases across the entire mass range, and for spin-independent cases across the high and low mass range.
}

\FullConference{%
  *** Particles and Nuclei International Conference - PANIC2021 ***\\
  *** 5 - 10 September, 2021 ***\\
  *** Online ***
}

\begin{document}
\maketitle

\vspace{-0.5cm}
\section{Introduction}
\vspace{-0.4cm}
Modern physics is striving to uncover the nature of Dark Matter (DM). Direct Detection (DD) experiments are at the forefront of this quest. The sensitivity of such searches has improved dramatically over the past few years, but they are still limited by practical factors such as the achievable mass of the detector targets and the recoil energy resolution threshold. As a consequence, it is natural to examine other systems in which DM interactions have observable effects.  Among the alternatives is to look at the effect of DM capture in stars~\cite{Gould:1987ju} and compact objects such as White Dwarfs (WD)~\cite{Bell:2021fye} and Neutron Stars (NS)~\cite{Bell:2018pkk, Bell:2019pyc, Bell:2020jou, Bell:2020lmm, Bell:2020obw, Anzuini:2021lnv}.  This work is focused on the latter.

DM's thermalisation and annihilation within these objects is one reason for much of this attention. This could result in visible heating~\cite{Bell:2018pkk, Bell:2019pyc, Bell:2021fye} of the objects. Additional consequences include the possibility of induced collapse into black holes when DM is non-annihilating or asymmetric, or modifications to the merger rate and its gravitational wave signature. 

Because of the extreme conditions present in NSs, the DM capture process differs significantly from that in the Sun. Due to this extreme environment, Gould's original formalism for DM capture in the Sun~\cite{Gould:1987ju} must be modified substantially. In recent works, a gradual introduction of these corrections has taken place.

In Ref.~\cite{Bell:2020jou}, a formalism was presented for neutron targets that consistently accounted for most of the physical effects.
These include: the NS internal structure, through solving the Tolman–Oppenheimer–Volkoff (TOV) equations coupled to an appropriate equation of state,  correct treatment of Pauli blocking for degenerate targets, relativistic kinematics, gravitational focusing, and multiple scattering effects for heavy DM. 

Ref. \cite{Bell:2020lmm} provided formalisms for lepton targets, electrons, and muons. 

In Ref.~\cite{Bell:2020obw, Anzuini:2021lnv}, we build on these improvements by focusing on the properties of baryonic targets (neutrons, i.e. protons and hyperons) in the interior of an extremely dense neutron star. In the NS interior, baryons experience strong interactions, and in DM-baryon collisions, the momentum transfer is large enough that baryons cannot be regarded as point particles. These features impact capture rate by up to three orders of magnitude for heavy NSs compared to less heavy NSs. 

\vspace{-0.5cm}
 \section{Neutron Star Heating}
 \vspace{-0.4cm}
During the course of their orbits around the center of the Galaxy, stars pass through large fluxes of DM particles.
Interacting with Standard Model particles within stars can result in DM loss of energy which can cause it to become gravitationally bound to the star if the loss of energy is large enough. As NSs are dense, the capture process is enhanced. 

In the case of DM scattering off SM particles within an NS core, if the cross section $\sigma$ of scattering is enough to capture all particles as they transit, $\sigma \gtrsim \sigma_{th}$, then the capture rate tends to the geometric limit~\cite{Bell:2018pkk},
\small
\begin{equation}
C_{\star} =  \frac{\pi R_{\star}^2 (1-B)}{v_\star B} \frac{\rho_\chi}{m_\chi} \erf\left(\sqrt{\frac{3}{2}}\frac{v_\star}{v_d}\right), \quad B = 1-\frac{2GM_{\star}}{c^2 R_{\star}},    
\label{eq:Cstar}
\end{equation}
\normalsize
where $\rho_\chi$ is the DM density, and $B$ is the value of space part of the metric on the surface of the star.
We have assumed a Maxwell-Boltzmann distribution for the DM speed, with $v_d$ the velocity dispersion and $v_\star$ the NS speed, which we assume to be comparable to the speed of the Sun.

Assuming that DM thermalizes then, after reaching the steady state, the energy contribution of each captured DM particle can be taken as the total initial energy $m_\chi(1/\sqrt{B_{core}}-1)$. Here $B_{core}$ is the value of $B$ in the core, and for the purpose of this paper it can be assumed to be a pure number in the range $0.2<B_{core}<0.5$, depending on the NS mass, with heavier NS giving smaller values. The DM contribution to the NS luminosity is
\small
\begin{equation}
L^{\infty,th}_{\rm DM} = m_\chi(1/\sqrt{B}-1) C_{\star} B^2= 4\pi \sigma_{SB} R_{core}^2 \left(T_{kin}^{\infty,th}\right)^4, 
\label{eq:lum}
\end{equation}
\normalsize
where $\sigma_{SB}$ is the Stefan-Boltzmann constant and $T^\infty=\sqrt{B}T$ is the temperature measured at large distance from the NS. Using eq.~\ref{eq:Cstar} and eq.~\ref{eq:lum}, we obtain~\cite{Bell:2018pkk}
\small
\begin{eqnarray}
T_{kin}^{\infty,th} &=& \left[ \frac{\rho_\chi (1-B)B}{4\sigma_{SB}v_\star }\left(\frac{1}{\sqrt{B_{core}}}-1\right)\erf\left(\sqrt{\frac{3}{2}}\frac{v_\star}{v_d}\right)\right]^{1/4}\nonumber \\
&\simeq& 1700 K \left(\frac{\rho_\chi}{0.4\GeV \cm^{-3}}\right)^{1/4}F\left(\frac{v_\star}{230\km \s^{-1}}\right), 
\label{eq:Tkin}
\end{eqnarray}
\normalsize
where
\small
\begin{equation}
F(x) = \left[\frac{\erf(x)}{x \erf(1)}\right]^{1/4}. 
\end{equation}
\normalsize
Note from Eq.~\ref{eq:Tkin} that a NS blackbody temperature of $T_{kin}^{\infty,th}\simeq1700$ K is expected in the case of maximal DM capture. 
Moreover, in the absence of another heating mechanism, this temperature will lead to radiation in the near infra-red, potentially detectable by the forthcoming JWST \cite{Baryakhtar:2017dbj}. 
It is important to remark that a cross section $\sigma_{i \chi}=\sigma_{th}$ maximises the kinetic heating; any larger cross section would produce the same effect as $\sigma_{i \chi}=\sigma_{th}$. For cross sections below $\sigma_{th}$, the capture rate is reduced as $C\propto \frac{\sigma_{i \chi}}{\sigma_{th}}$ and the kinetic heating temperature decreases as  $(\frac{\sigma_{i \chi}}{\sigma_{th}})^{1/4}$. If one allows DM to annihilate in the center of the NS, the equilibrium temperature is raised to
\small
\begin{eqnarray}
T_{ann}^{\infty,th} &=& \left(\frac{1}{1-\sqrt{B_{core}}}\right)^{1/4} T_{kin}^{\infty,th} \simeq 2300 K \left(\frac{\rho_\chi}{0.4\GeV \cm^{-3}}\right)^{1/4}F\left(\frac{v_\star}{230\km \s^{-1}}\right), 
\label{eq:Tann}
\end{eqnarray}
\normalsize

\begin{figure}
\vspace{-0.5cm}
\centering
\includegraphics[width=0.4\textwidth]{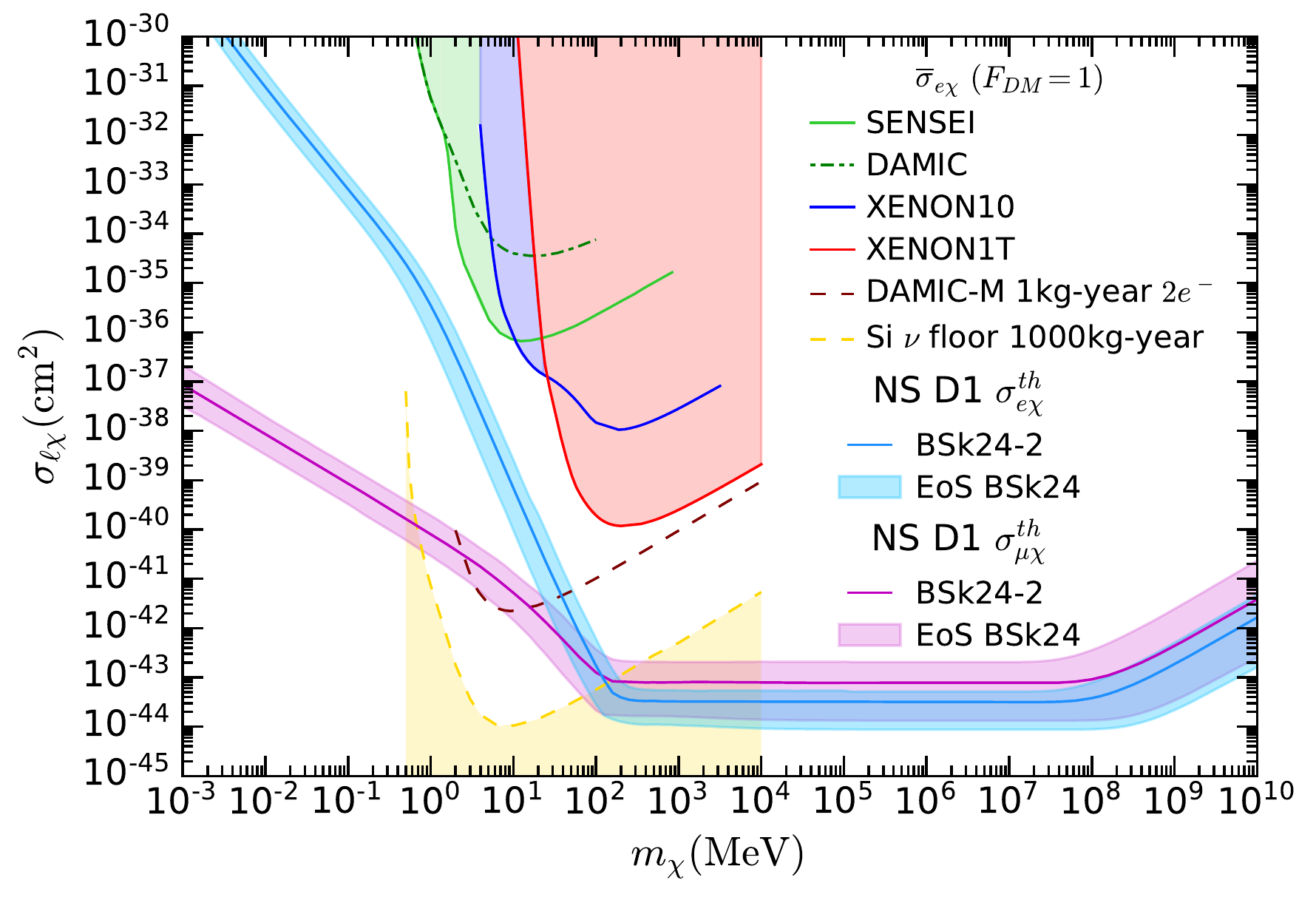}
\includegraphics[width=0.4\textwidth]{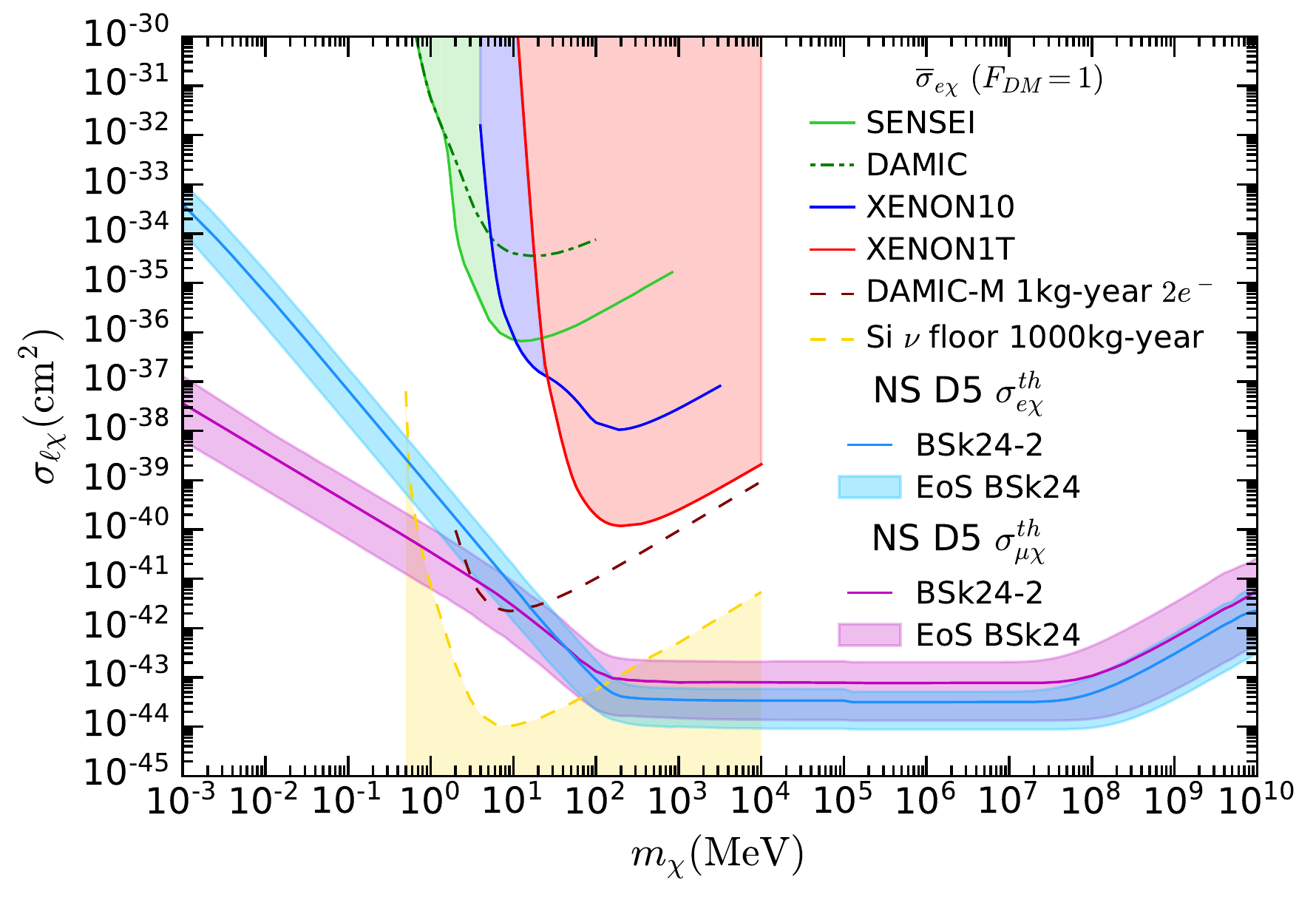}
\vspace{-0.5cm}
\caption{\small DM-lepton threshold cross section for operators D1 (left) and D5 (right) for the EoS BSk24, presented in \cite{Bell:2020lmm}. The solid blue (electron) and magenta (muon) lines represent $\sigmath$, computed assuming a $1.5M_\odot$ NS mass, while the shaded bands represent the expected range due to variation of the NS mass between $1M_\odot$ and $2M_\odot$. For comparison we show leading electron recoil bounds for heavy mediators from SENSEI, DAMIC, Xenon10, Xenon1T, projected sensitivities from DAMIC-M as well as the neutrino floor for silicon detectors. }
    \label{fig:sigmathe}
    \vspace{-0.5cm}
\end{figure}

For a given DM model, setting $T_{eq}=T_{kin}^{\infty,th}$ if DM is not allowed to annihilate or $T_{eq}=T_{ann}^{\infty,th}$ if it is, one can use the observation of a cold old NS of temperature $T$ to constrain the DM interaction cross section as

\small
 \vspace{-0.5cm}
\begin{equation}
    \sigma \lesssim \sigma_{th} \left(\frac{T}{T_{eq}}\right)^4.\label{eq:constraint}
\end{equation}
\normalsize

In the next section will therefore be devoted to the accurate computation of the Capture rates $C$ that will allow us to infer a precise value of $\sigma_{th}$, necessary for an accurate result of Eq. \ref{eq:constraint}.

\vspace{-0.5cm}
\section{Dark Matter Capture}
\vspace{-0.3cm}

The Capture rate $C$ can be calculated starting from the interaction rate $\Omega^{-}$ \cite{Bell:2020jou,Bell:2020lmm,Anzuini:2021lnv}

\small
 \vspace{-0.5cm}
\begin{eqnarray}
C &=& \frac{4\pi}{\vstar} \frac{\rho_\chi}{m_\chi} {\rm Erf }\left(\sqrt{\frac{3}{2}}\frac{\vstar}{v_d}\right)\int_0^{\Rstar}  r^2 \frac{\sqrt{1-B(r)}}{B(r)} \Omega^{-}(r)  \, dr, \label{eq:captureM2} 
\\
\Omega^{-}(r) &=& \frac{\zeta(r)}{32\pi^3}\int dt dE_n ds  \frac{|\overline{M}(s,t)|^2}{s^2-(m_n^2-m_\chi^2)^2} \frac{E_n}{m_\chi}\sqrt{\frac{B(r)}{1-B(r)}}
\frac{s}{\gamma(s)}
\fFD(E_n,r)[1-\fFD(E_n^{'},r)],\label{eq:intrateideal}
\end{eqnarray}
\normalsize
where $B(r)$ is the radial-dependent value of the metric, $\fFD$ is the Fermi-Dirac distribution, $|\overline{M}(s,t)|^2$ is the squared and spin averaged matrix element for the interaction that one wants to consider, $E_n^{'}$ is the final energy of the target particle (that can be obtained by kinematics as a function of the other variables), and $\gamma(s),\zeta(r)$ are defined in Ref.\citep{Bell:2020jou, Bell:2020lmm}\footnote{The factor $\zeta$ can be taken to be one when one considers the interactive approach, see \cite{Anzuini:2021lnv}.}. 

\begin{figure}
\vspace{-0.5cm}
    \centering 
\includegraphics[width=0.48\textwidth]{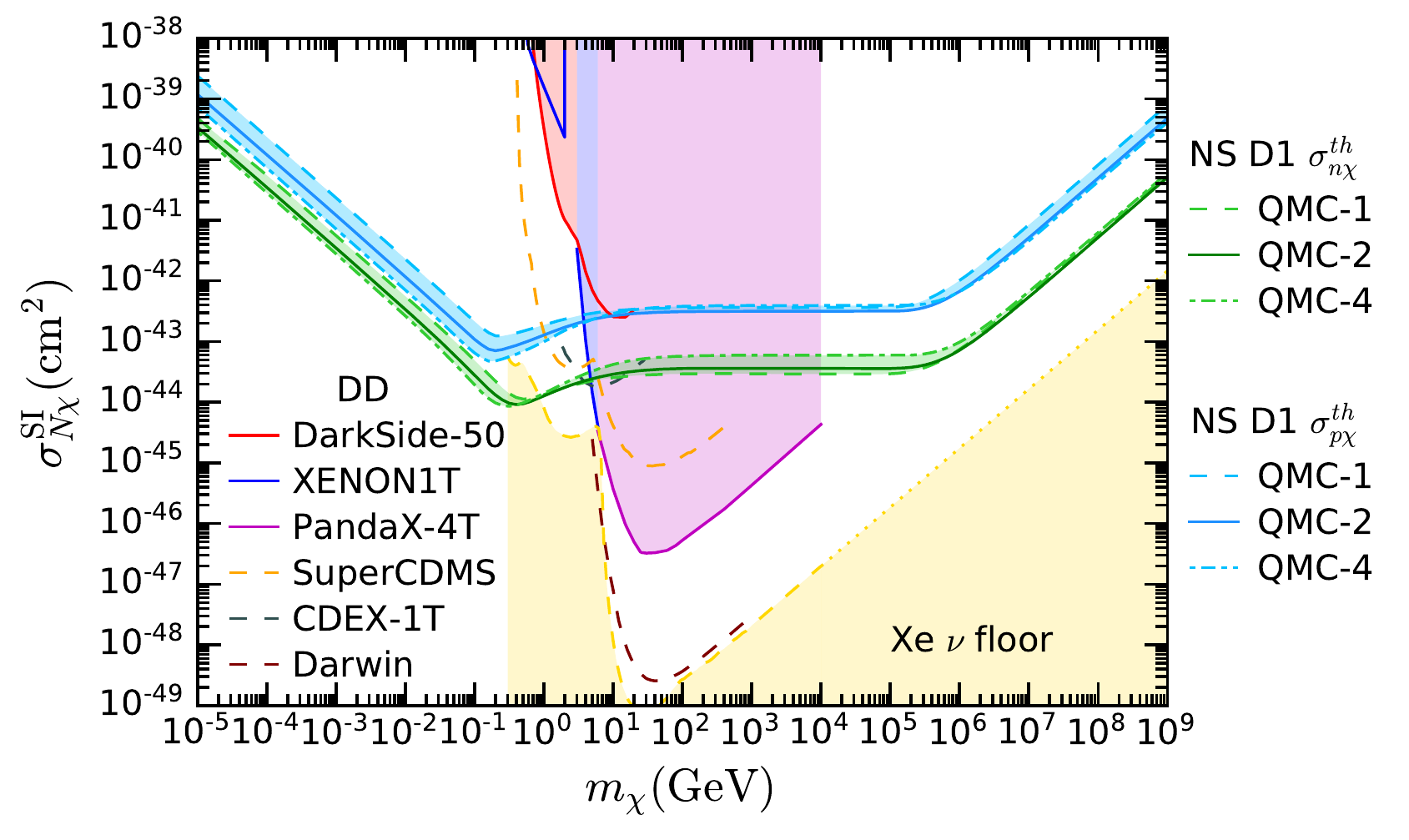}
\includegraphics[width=0.8\textwidth]{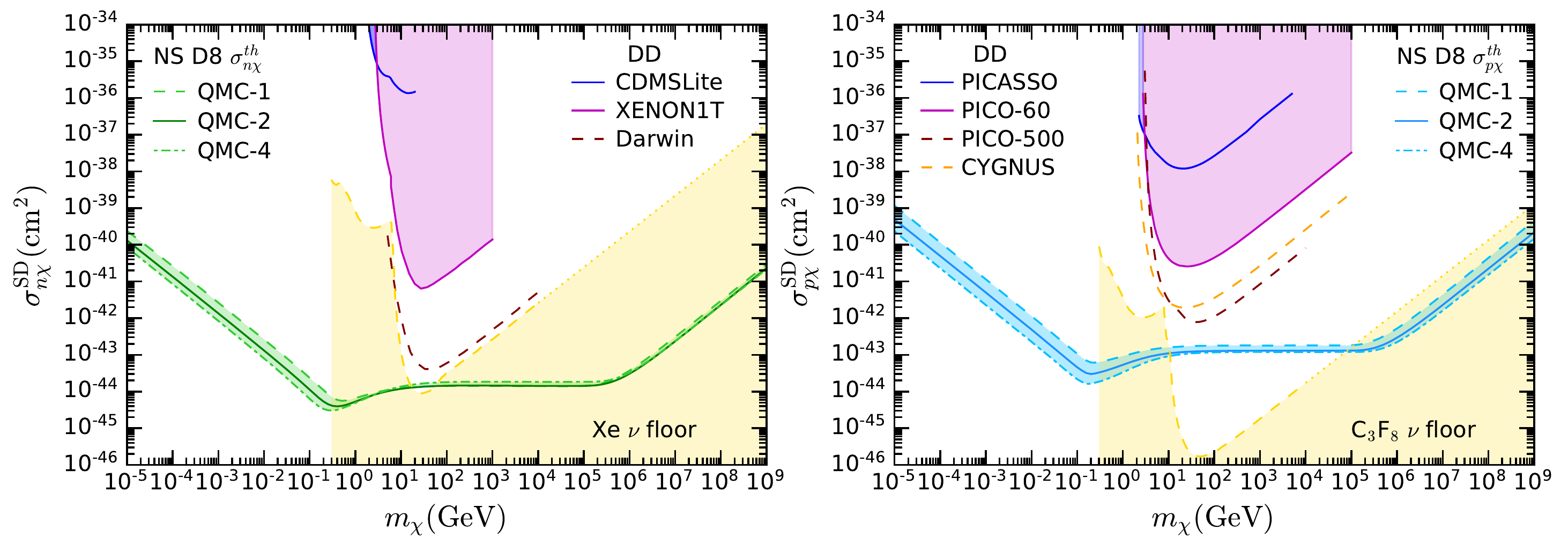}
\vspace{-0.5cm}
    \caption{\small DM-nucleon threshold cross section for operators D1 (top) and D8 (bottom) for the QMC EoS family, presented in \cite{Anzuini:2021lnv}. The solid line represents $\sigmath$, computed assuming the NS configuration QMC-2. 
    We also show for comparison the leading spin-independent (SI) and spin-dependent (SD) DD limits from CDMSLite, DarkSide-50,  Xenon1T, PandaX-4T, PICASSO and PICO-60,  projected sensitivities from SuperCDMS SNOLAB Ge/Si, CDEX-1T, CYGNUS  $10\m^3$,  PICO-500 and Darwin, as well as the neutrino coherent scattering background for xenon and $\mathrm{C_3F_8}$ bubble chamber detectors.  
    } 
    \label{fig:sigmath}
    \vspace{-0.5cm}
\end{figure}

In the small mass range, $m_\chi \lesssim 1\GeV$, one can see that the Pauli Blocking term $[1-\fFD(E_n^{'},r)]$ in  Eq. \ref{eq:captureM2} effectively introduces a suppression that is of $\mathcal{O}\left(\frac{m_\chi}{p_F}\right)^{1+n}$, where $p_F$ is the Fermi momentum and is $p_F\sim 1\GeV$ for NS, and $n$ is the power of the momentum transfer $-t=Q^2$ that the squared matrix element is proportional to. For standard interactions, $n=0$. 
In the large mass range, $m_\chi\gtrsim 10^5\GeV$, to account for multiple scattering, one needs to include an additional factor in Eq. \ref{eq:captureM2} (see Ref. \cite{Bell:2020jou,Bell:2020lmm,Anzuini:2021lnv}), while in the intermediate mass range, $1\GeV\lesssim m_\chi\lesssim 10^5\GeV$, this equation can be simplified by removing the Pauli Blocking term $[1-\fFD(E_n^{'},r)]$ and performing some integration steps analytically, see Ref. \cite{Bell:2020jou,Bell:2020lmm,Anzuini:2021lnv}. Eq. \ref{eq:captureM2} can be used to compute capture rates also in WD for electron targets \cite{Bell:2021fye}.


\section{Results}
\vspace{-0.3cm}

We use Eq. \ref{eq:captureM2} to calculate Capture Rates for various EFT operators listed in \cite{Bell:2020jou,Bell:2020lmm,Anzuini:2021lnv}. We report in Fig. \ref{fig:sigmathe} and Fig. \ref{fig:sigmath} the resulting threshold cross sections for leptons and nucleons respectively, comparing them to bounds from existing experiments. We can notice that, apart the case of SI interactions with nucleons and $m_\chi \gtrsim 10\GeV$, all
the limits for existing experiments are orders of magnitude weaker than the expected NS reach. Only future DD experiments such as DAMIC-M could be competitive, for leptons, and, even then, only in a narrow mass range $3 \MeV\lesssim m_\chi\lesssim  30\MeV$. Both for leptons and nucleons, we can notice that in some part of the mass spectrum, NS are actually able to probe cross sections that are below the DD neutrino floor.

\vspace{-0.5cm}
\section{Summary}
\vspace{-0.3cm}
We have seen that NS can be cosmic laboratories to probe DM scattering interactions. By interacting with DM, NS can be very efficient in capturing large quantities of DM particles. These interaction are however also a source of heating, and in the absence of other heating mechanisms, they set an equilibrium temperature, estimated to be between $1700K$ and $2300K$, depending on weather DM can annihilate or not. 
DM scatterings in NS have however kinematic conditions completely different from Earth. The large energy and momentum transfers require a different treatment comparing to Capture of DM in the Sun or Direct Detection. A precise result of the capture rates is necessary to estimate the threshold cross sections. The values of such cross sections can be used to place very stringent bounds on DM interactions in case of the observation of an old and cold NS, that are in most cases orders of magnitude stronger than existing and future DD experiments, and sometimes even below the neutrino background.

\vspace{-0.5cm}
\label{Bibliography}
\bibliography{Bibliography} 


\end{document}